\documentclass[apj]{emulateapj}
\usepackage{times}
\usepackage[normalem]{ulem}
\usepackage{epsfig}
\usepackage{natbib}
\usepackage{amsfonts}
\usepackage{amsmath}
\usepackage{multirow}
\usepackage{enumerate}
\usepackage[usenames]{color}
\usepackage[plainpages=false, colorlinks=true, anchorcolor=blue, linkcolor=blue, citecolor=blue, bookmarks=false]{hyperref}
\newcommand{\jcap}{JCAP}

\newcommand{\be}{\begin{eqnarray}}
\newcommand{\msun}{M_\odot}
\newcommand{\ee}{\end{eqnarray}}
\newcommand{\lp}{\left(}
\newcommand{\rp}{\right)}
\newcommand{\lb}{\left[}
\newcommand{\rb}{\right]}

\begin{document}


\slugcomment{Accepted for publication in The Astrophysical Journal}

\title{Ultrahigh-Energy Cosmic Rays from the ``En Caul" Birth\footnotemark[1]  of Magnetars}

\author{Anthony L. Piro\altaffilmark{2} and Juna A. Kollmeier\altaffilmark{2,3}}


\footnotetext[1]{The birth ``en caul'' refers to the rare circumstance in which a newborn emerges in a fully intact amniotic sac.  Birth of this nature is considered a sign of good fortune in many cultures.  Here, we refer to the newborn magnetar similarly surrounded by a small amount of natal material and similarly fortunate as a cosmic-ray accelerator.}

\altaffiltext{2}{Carnegie Observatories, 813 Santa Barbara Street, Pasadena, CA 91101, USA; piro@obs.carnegiescience.edu}

\altaffiltext{3}{School of Natural Sciences, Institute for Advanced Study, Einstein Drive, Princeton, NJ 08540, USA}


\begin{abstract}
Rapidly-spinning magnetars can potentially form by the accretion induced collapse of a white dwarf or by neutron star mergers if the equation of state of nuclear density matter is such that two low mass neutron stars can sometimes form a massive neutron star rather than a black hole.
In either case, the newly born magnetar is an attractive site for producing ultrahigh-energy cosmic rays (particles with individual energies exceeding $10^{18}\,{\rm eV}$; UHECRs). The short-period spin and strong magnetic field are able to accelerate particles up to the appropriate energies, and the composition of material on and around the magnetar may naturally explain recent inferences of heavy elements in UHECRs. We explore whether the small amount of natal debris surrounding these magnetars allows the UHECRs to easily escape. We also investigate the impact on the UHECRs of the unique environment around the magnetar, which consists of a bubble of relativistic particles and magnetic field within the debris. Rates and energetics of UHECRs are consistent with such an origin even though the rates of events that produce rapidly-spinning magnetars remain very uncertain. The low ejecta mass also helps limit the high-energy neutrino background associated with this scenario to be below current IceCube constraints over most of the magnetar parameter space. A unique prediction is that UHECRs may be generated in old stellar environments without strong star formation in contrast to what would be expected for other UHECR scenarios, such as active galactic nuclei or long gamma-ray bursts.
\end{abstract}



\section{Introduction}

The origin of ultrahigh-energy cosmic rays (UHECRs), particles that can have a kinetic energy as large as a thrown baseball, remains one of the outstanding questions in astrophysics \citep[e.g,][]{Kotera11,Letessier11}. Their mystery has sparked a wide range of models for the acceleration sites of UHECRs, with varying levels of success in explaining their energetics and spectrum. This includes, but is not limited to, galaxy clusters \citep{Inoue07,Kotera09}, active galactic nuclei \citep[AGNs;][]{Dermer09,Takami11}, classical gamma-ray bursts \citep[GRBs; ][]{waxman04,Waxman06,Dermer10}, low-luminosity GRBs \citep{Murase06}, magnetically dominated jets in the millisecond protomagnetar model for GRBs \citep{Metzger11}, engine-driven supernovae \citep{Wang07,Chakraborti11}, AGN flares \citep{Farrar09}, and newly-born, fast-rotating pulsars \citep{Fang12,Kotera15}.

Observationally, there are many clues about UHECRs to help constrain these models. The flux suppression at the highest energies \citep{Abbasi08,Abraham08} is consistent with UHECRs interacting with cosmic microwave background photons or other sources of extragalactic light, known as the GZK effect \citep{Greisen66,Zatsepin66}. This strongly suggests an extragalactic origin, but also limits their source to be in our local universe within $\sim100\,{\rm Mpc}$.  The energy injected into \mbox{UHECRs} is \mbox{$\sim10^{43.5}-10^{44}\,{\rm erg\,Mpc^{-3}\,yr^{-1}}$} \citep{Katz09}, which must be satisfied by the combined rate and energy production of any model. Most recently, the Auger Observatory results point to a composition of UHECRs that may be dominated by heavier nuclei at the highest energies \citep{Aab14a,Aab14b}. If confirmed, this would pose a problem for any model that produces UHECRs simply as protons.

At the same time, motivated by different physical considerations, there has been increasing focus on the possibility of high-energy events that may result in a quickly-spinning magnetar with a {\it small} amount of surrounding material. These mainly come in three varieties, the first of which is the accretion induced collapse (AIC) of a white dwarf (WD) to a neutron star (NS). In this case, either through the merger of two carbon-oxygen WDs, or accretion onto an oxygen-neon-magnesium WD, the WD electron captures, loses its pressure support and collapses to an NS \citep{Canal76,Nomoto91}. The large angular momentum associated with such an event may naturally lead to short-period spins and a strong magnetic field \citep{Duncan92,Price06,Zrake13}. In the second case, the merger or grazing collision of two NSs may in some circumstances \citep[depending on their masses and the maximum mass allowed by the equation of state of nuclear density matter;][]{Hebeler13} produce a stable, high-mass NS rather than a black hole \citep[e.g.,][]{Ozel10,Giacomazzo13,Kiziltan13}. In the third case, the massive NS may be present, but only for a limited time due to support by thermal pressure and/or differential rotation \citep{Lehner12,Paschalidis12,Kaplan14}. In any case, there is considerable interest in the potential transient signature that these events may make \citep{Piro13,Metzger14,Piro14} as well as the exciting possibility that they may be observed as electromagnetic counterparts to gravitational wave sources.

Here we propose that such events may also be ideal sites for producing UHECRs. The fast spin and high magnetic field of a newly born magnetar provide a natural location for accelerating UHECRs to the correct energies. Most critically, the low amount of natal material surrounding the magnetars should, in principle, make it easier for UHECRs to escape the magnetar and seed the local universe. In this work, we examine this possibility to determine whether this provides a viable site for UHECR production.  In particular, we investigate whether the unique environment around the magnetar, which includes a bubble of relativistic particles and magnetic field inflated by the magnetar \citep{Metzger14,Siegel15a,Siegel15b}, impedes and/or imprints itself on the UHECRs in any way. Furthermore, we investigate whether such a scenario will have implications for the rates and environments where the UHECRs may come from. 

Our model differs in several critical respects from previous studies that focused on fast-spinning pulsars \citep[e.g.][]{Arons03, Fang12}.  In particular, these works have focused on a core-collapse scenario as a precursor to the UHECR engine.  As a result, a large amount of ejecta material that surrounds the pulsar (the pulsar wind nebula; usually assumed to be $\sim 5-10 \msun$ in line with estimates from the Crab nebula) in comparison to the ``en caul'' birth we consider with only $\sim0.1\,M_\odot$ of surrounding material. A large amount of ejecta generically makes it more difficult for UHECRs to escape the core-collapse environment as shown by \citet{Fang12}.  Another important difference is that we focus on much earlier times after the magnetar is born ($\sim$~days rather than $\sim$~months to years), when the nebula around the pulsar is denser in electron/positron pairs. The density of charged particles is then better described in a saturated regime \citep{Svensson87} rather than simply given by the Goldreich-Julian density \citep{Goldreich69}. We discuss the impact of these differences on our model below.

In Section \ref{sec:model}, we present a simplified model for following the time-dependent features of the environment around the newly created magnetar. In Section \ref{sec:results}, we use these models to calculate our main results for what energy particles may come from the magnetar and how these particles propagate through the environment of the magnetar. In Section \ref{sec:rates}, we estimate the expected rates of such events in comparison to measurements of UHECRs. We also discuss various constraints on our scenario, including the rate of expected associated electromagnetic transients, galactic and intergalactic propagation issues, and the high-energy neutrino background. Finally, in Section \ref{sec:conclusions}, we summarize our results and discuss potential future work.

\section{The Magnetar and Surrounding Region}
\label{sec:model}

To follow the environment around the newly born magnetar, we construct a toy model that replicates the key physical features. For more detailed discussions and calculations, with a particular focus on the electromagnetic signatures, the reader should consult \citet{Metzger14} and \citet{Siegel15a,Siegel15b}.

The general picture we are considering is that, whether born from an AIC or from an NS merger, the initial state is a magnetar with surface dipole field $B_*\sim10^{14}-10^{15}\,{\rm G}$ surrounded by $M_{\rm ej}\sim0.01\,M_\odot$ of ejecta with a velocity of $v_{\mathrm{ej},i}\sim0.1c$ \citep{Dessart09,Metzger09a,Metzger09b,Lee09,Fernandez13}. The specific values can of course change and we will explore the dependencies on them. Although the magnetar may be initially spinning near its centrifugal break-up limits, torques from a neutrino-driven wind \citep{Thompson04} or gravitational wave losses \citep{Piro11} likely spin it down during the first milliseconds to spin periods of $P_i=2\pi/\Omega_i\sim2-5\,{\rm ms}$, and thus we consider this the ``initial'' period for purposes of our calculations.

\subsection{Energy Input by the Magnetar}

The initial rotational energy of the magnetar with moment of inertia $I$ is
\be
	E_{\rm rot} = \frac{1}{2}I\Omega_i^2 =  4.9\times10^{51}I_{45}\lp\frac{P_i}{2\,{\rm ms}} \rp^{-2}{\rm erg},
	\label{eq:erot}
\ee
where $I_{45}=I/10^{45}\,{\rm g\,cm^2}$. The spin down of the magnetar with magnetic moment $\mu=R_*^3B_*$ injects energy with a time-dependent rate of
\be
	L_{\rm sd}(t) = \frac{\mu^2\Omega(t)^4}{c^3}= \frac{\mu^2\Omega_i^4}{c^3}\lp1+\frac{t}{t_{\rm sd}} \rp^{-2},
\ee
where the spin down timescale is
\be
	t_{\rm sd} = \frac{Ic^3}{2\mu^2\Omega_i^2} = 22.8\mu_{33}^{-2}I_{45}\lp \frac{P_i}{2\,{\rm ms}} \rp^2{\rm min}.
\ee
The ejecta around the magnetar is inflated by this energy input with a time-dependent velocity of
\be
	v_{\rm ej}(t) = \frac{dR_{\rm ej}}{dt} = \lb 2M_{\rm ej}^{-1}\int_0^tL_{\rm sd}(t')dt' + v_{\mathrm{ej},i} ^2 \rb^{1/2},
\ee
which can be simply integrated to find $R_{\rm ej}(t)$. At late times the velocity of the ejecta can get high
\be
	v_{\rm ej}(t\gg t_{\rm sd}) &\approx& \lp 2E_{\rm rot}/M_{\rm ej} \rp^{1/2}
	\nonumber
	\\
		&\approx& 0.7c I_{45}^{1/2}M_{-2}^{-1/2}\lp\frac{P_i}{2\,{\rm ms}} \rp^{-1},
\ee
where $M_{-2}=M_{\rm ej}/10^{-2}\,M_\odot$. To simplify our analysis, we focus on values of $E_{\rm rot}$ and $M_{\rm ej}$ where $v_{\rm ej}\lesssim c$.

\subsection{Evolution of the Perimagnetar Environment}

In order to elucidate the UHECR propagation, we divide the perimagnetar environment into two primary regions: (1) the ejecta shell and (2) the nebula (or bubble) of relativistic particles around the magnetar \citep[see the diagram in][]{Metzger14}. In each of these regions, both the particle density and radiation content must be followed since these can interact with UHECRs and alter their fate. In this simplified picture, we do not explicitly follow the dynamics of the termination shock between the magnetar wind and the nebula. Nevertheless, this shock is important for converting the initial Poynting-flux dominated outflow of the magnetar into kinetic energy, as has been extensively studied in past work \citep[e.g.,][]{Lyubarsky01,Arons08,Lyubarsky10,Granot11,Porth13}. Since the presence of the termination shock does not impact the shell dynamics beyond providing a potential site for particle acceleration, we mostly ignore it when solving for the properties of the nebula.

For the ejecta region, the density simply evolves as the ejecta expands as
\be
	\rho_{\rm ej}(t) = \frac{3M_{\rm ej}}{4\pi R_{\rm ej}(t)^3}.
\ee
The composition is thought to be primarily iron-rich elements due to the higher electron fraction produced by neutrino irradiation from the magnetar \citep{Metzger09b,Fernandez13}.  We estimate the number density of nuclei in the ejecta as $n_{\rm ej}\approx \rho_{\rm ej}/56m_p$, where $m_p$ is the proton mass.  

The second region is the relativistic bubble interior to the ejecta. This is inflated by the magnetar spin down luminosity, which can be partitioned into a combination of e$^\pm$ pairs, magnetic field, radiation, and, importantly for the study here, UHECRs. The high density of radiation and magnetic field within the nebula gives rise to copious e$^\pm$ pairs, with the strength of the pair creation described by the ``compactness parameter''
\be
	\ell = \frac{3E_{\rm nth}\sigma_{\rm Th}}{4\pi R_{\rm ej}^2m_ec^2}
\ee
where $E_{\rm nth}$ is the energy in non-thermal radiation within the nebula (which we describe in more detail below) and $\sigma_{\rm Th}$ is the Thomson cross section. When $\ell\gtrsim1$, the number density of these pairs can be estimated by balancing the rate of pair creation and annihilation \citep{MetzgerVurm14,Metzger14}, resulting in
\be
	n_\pm (t)= \lb  \frac{4YL_{\rm sd}(t)}{\pi R_{\rm ej}(t)^3\sigma_{\rm Th}m_ec^3}  \rb^{1/2},
	\label{eq:pairs}
\ee
where $\sigma_{\rm Th}=6.65\times10^{-25}\,{\rm cm}$ is the Thomson cross section and $Y\approx0.1$ is the pair multiplicity factor for a saturated cascade \citep[essentially the fraction of the magnetar spin down that goes into pairs;][]{Svensson87}. In our case, $\ell\gtrsim1$ for the times we consider as we show explicitly below.  This density is in contrast to other work on UHECRs from magnetars in a core-collapse scenario \citep{Fang12,Lemoine15} that use a Goldreich-Julian density instead \citep{Goldreich69}. Another difference is that Equation (\ref{eq:pairs}) assumes that the e$^\pm$ are fairly sub-relativistic. This is reasonable because even if the pairs are created with a high Lorentz factor, the ratio of the Compton cooling timescale $t_{\rm C}$ to the time the nebula has been expanding $t$ is
\be
	\frac{t_{\rm C}}{t} &=& \frac{\pi m_ecR_{\rm ej}^3}{\sigma_{\rm Th}E_{\rm nth}t}
	\sim \frac{\pi m_ec v_{\rm ej}^3t}{\sigma_{\rm Th}L_{\rm sd}}
	\nonumber
	\\
	&\sim& 0.1 \mu_{33}^{2}I_{45}^{-1/2}M_{-2}^{-3/2}\lp\frac{t}{1\,{\rm day}} \rp^{-2},
\ee
and thus the pairs cool rapidly in the dense nebula.

In addition to e$^\pm$ pairs, the nebula will be filled with magnetic field. As mentioned above, the general picture is that the magnetar outflow is Poynting-flux dominated inside of the termination shock, but dissipation near the shock results in a much smaller mean field in the rest of the nebula. Since the exact conversion factor is not known, especially for the extreme case of magnetars we consider here, we quantify this uncertainty with a factor $\eta_B$ that is the fraction of the magnetar spin down that goes into this magnetic field. This then gives
\be
	B_{\rm neb}(t) \approx \lp\frac{6\eta_B}{R_{\rm ej}(t)^3} \int_0^tL_{\rm sd}(t')dt'  \rp^{1/2},
	\label{eq:bneb}
\ee
as an estimate of the time dependent mean magnetic field in the nebula. We note that this magnetic field was left out of the energy budget of the nebula in the calculations by \citet{Metzger14}, but this only represents a small correction since $\eta_B\lesssim0.1$.

Next we consider is the radiation content in both the nebula and surrounding ejecta. These are highly coupled. The non-thermal radiation in the nebula is dominated by synchrotron and inverse-Compton emission \citep{MetzgerVurm14,Murase15}. This is then partially absorbed and thermalized by bound-free reactions by the iron-peak elements in the ejecta. In the Appendix of \citet{Metzger14}, we show that this interplay can be accurately followed in each region by simply considering two coupled differential equations. The non-thermal energy of the nebula, $E_{\rm nth}$, is simply given by the energy injection from spin down minus adiabatic and absorptive losses as a function of time
\be
	\frac{dE_{\rm nth}}{dt} = L_{\rm sd} - L_{\rm abs} - \frac{E_{\rm nth}}{t+t_i}.
	\label{eq:enth}
\ee
What is absorbed by the ejecta from the non-thermal reservoir is converted to heat and therefore the evolution of the thermal radiation of the ejecta, $E_{\rm th}$, evolves as
\be
	\frac{dE_{\rm th}}{dt} = L_{\rm abs} - L_{\rm rad} - \frac{E_{\rm th}}{t+t_i}.
	\label{eq:eth}
\ee
These equations have a number of terms that we now describe in detail. The rate of absorption of non-thermal radiation is
\be
	L_{\rm abs} = \frac{f_{\rm abs}E_{\rm nth}}{t_{\rm diff,neb}},
\ee
where $f_{\rm abs}$ is the fraction of non-thermal radiation absorbed by the ejecta and the diffusion time across the nebula is
\be
	t_{\rm diff,neb} = \lp \sigma_{\rm Th}R_{\rm ej}n_\pm + 1 \rp R_{\rm ej}/c.
\ee
The term $L_{\rm abs}$ appears in both Equations (\ref{eq:enth}) and (\ref{eq:eth}) as it mediates the transport of energy between these two reservoirs of radiation. The rate of radiative energy loss from the ejecta is simply
\be
	L_{\rm rad} =  \frac{E_{\rm th}}{t_{\rm diff,ej}},
\ee
where
\be
	t_{\rm diff,ej} = \lp \frac{\kappa_{\rm es}M_{\rm ej}}{4\pi R_{\rm ej}^2} + 1 \rp \frac{R_{\rm ej}}{c},
\ee
is the photon diffusion time through the ejecta and $\kappa_{\rm es}$ is the electron scattering opacity of the ejecta. The term $L_{\rm rad}$ is important for estimating the thermal radiation observed from the heated ejecta. The timescale $t_i=R_{\mathrm{ej},i}/v_{\mathrm{ej},i}$ is the initial expansion timescale of the ejecta. It is included to produce well-behaved solutions at early times and the results are largely insensitive to its value. Finally, the last term in each expression accounts for the adiabatic expansion of the relativistic gas, and hence has the familiar scaling of $E_{\rm (n)th}/t$ at late times.

\begin{figure}
\epsscale{1.2}
\plotone{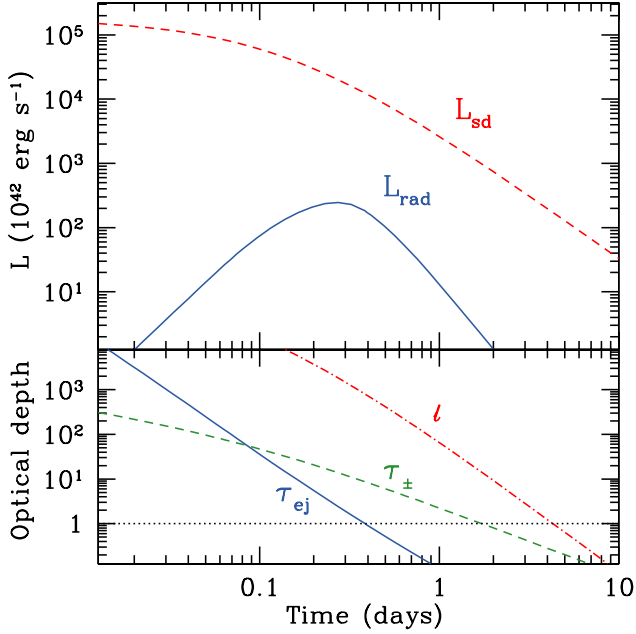}
\caption{Example evolution of the ejecta and nebula around a newly born, quickly rotating magnetar using $P_i=3\,{\rm ms}$, $B_*=5\times10^{14}\,{\rm G}$, and $M_{\rm ej}=0.01\,M_\odot$. The top panel shows the energy input by the spin down $L_{\rm sd}$ (red, dashed line) and the thermal radiation leaving the ejecta $L_{\rm rad}$ (blue, solid line). The bottom panel shows the electron scattering optical depths for the pairs in the nebula (green, dashed line) and the ejecta (blue, solid line), $\tau_\pm$ and $\tau_{\rm ej}$, respectively, as well as the compactness $\ell$ (red, dot-dashed line). The horizontal, black dotted line simply indicates where these quantities become unity.}
\label{fig:properties}
\epsscale{1.0}
\end{figure}

\begin{figure}
\epsscale{1.2}
\plotone{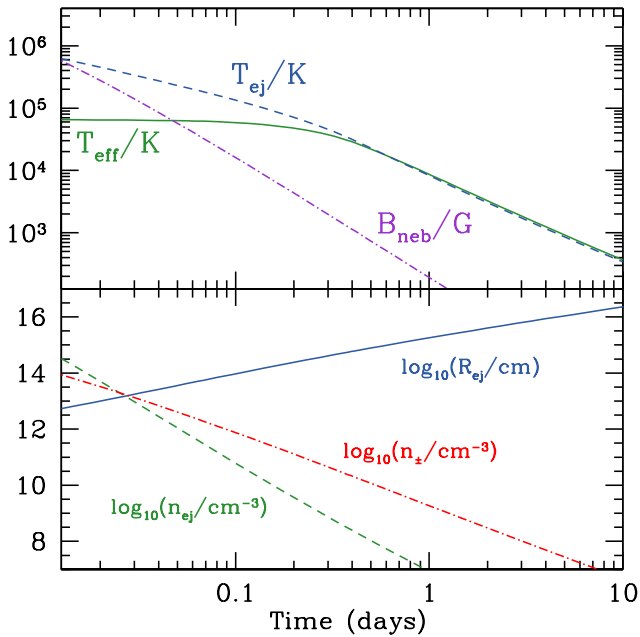}
\caption{Further properties of the ejecta and nebula from the model in Figure \ref{fig:properties}. The top panel shows the temperature of the ejecta $T_{\rm ej}$ (blue, dashed line), the effective temperature of the ejecta $T_{\rm eft}$ (solid, green line), and the mean magnetic field in the nebula $B_{\rm neb}$ using $\eta_B=0.1$ (dot-dashed, purple line). The bottom panel shows the density pairs in the nebula and the ejecta, $n_\pm$ (red, dot-dashed line), the density of ions in the ejecta $n_{\rm ej}$ (green, dashed line), and the radius of the ejecta $R_{\rm ej}$ (blue, solid line). The convergence of lines in the bottom panel is just a coincidence for this particular model.}
\label{fig:properties2}
\epsscale{1.0}
\end{figure}

From the above description, given specific properties for the magnetar and ejecta, one can solve for the evolution in $n_{\rm ej}$, $n_\pm$, $E_{\rm nth}$ and $E_{\rm th}$ to investigate the possible UHECRs evolution. An example solution is presented in Figures \ref{fig:properties} and \ref{fig:properties2} for $P_i=3\,{\rm ms}$, $B_*=5\times10^{14}\,{\rm G}$, and $M_{\rm ej}=0.01\,M_\odot$. In this case we use $f=1$, which generally reproduce the features of more detailed treatments \citep{Metzger14}. Starting with Figure \ref{fig:properties}, the thermal light curve $L_{\rm rad}$ shows the characteristic supernova-like shape, which peaks on a $\sim1\,{\rm day}$ timescale, potentially being the best electromagnetic signature for identifying such events \citep[note the non-thermal X-rays have a similar luminosity and time scale as well,][]{Metzger14}. The observed $L_{\rm rad}$ is well below $L_{\rm sd}$ because of the large optical depth in pairs hindering the transfer of energy from the nebula to the ejecta, as can be seen from $\tau_\pm\gtrsim \tau_{\rm ej}$ near the light curve peak. Here we use $\tau_\pm=\sigma_{\rm Th}R_{\rm ej}n_\pm$ and $\tau_{\rm ej}=\kappa_{\rm es}M_{\rm ej}/4\pi R_{\rm ej}^2$. In the bottom panel, we plot $n_{\rm ej}$ and $n_\pm$, which will be important for understanding UHECR propagation in subsequent sections. We also plot the compactness $\ell$ in Figure \ref{fig:properties}, demonstrating that it is greater than unity for most of the evolution, justifying our treatment of the e$^\pm$ pairs in the saturated regime.

In Figure \ref{fig:properties2}, we summarize more important quantities from the evolution of this model. The top panel shows both the temperature of the ejecta along with the effective temperature of the thermal optical signature. These become roughly equal once $\tau_{\rm ej}\lesssim1$. The mean magnetic field in the nebula $B_{\rm neb}$, calculated from Equation (\ref{eq:bneb}), will be important for understanding the acceleration of UHECRs as well as potentially cooling them via synchrotron radiation. In the bottom panel we summaries the critical densities in each regime, along with the radius of the expanding ejecta $R_{\rm ej}$.

\section{UHECR Energies and Propagation Estimates}
\label{sec:results}

In Section \ref{sec:model}, we presented the basic physical model and demonstrated that we can follow the evolution of the en caul magnetar in a straightforward fashion.  Here, we estimate the energy of the individual UHECR particles and show how the perimagnetar environment affects the ultimate destiny of UHECRs produced from this channel. 

\subsection{Maximum Energy of UHECRs}

We first summarize the basic energetics to show that, indeed, the magnetar surface can easily accelerate CRs to ultra-high energies. Relativistic magnetic rotators have voltage drops across their magnetic field of
\be
	\Phi(t) = \frac{\mu\Omega(t)^2}{c^2}.
\ee
A particle of charge $Z$ that is stripped off the surface of the magnetar (by some combination of strong electric fields, bombardment of particles, and boiling by stellar heat) can be accelerated to an energy
\be
	E_{\rm surf} = Ze\Phi = 3.3\times10^{21}Z\mu_{33}\lp\frac{P}{2\,{\rm ms}} \rp^{-2}\,{\rm eV}.
	\label{eq:emax}
\ee
This shows that in practice it is reasonable to expect nuclei accelerated to sufficient energies to produce UHECRs, especially for large charges ($Z>1$). The problem is that when this acceleration occurs within the light cylinder of the magnetar $R_{\rm LC}=\Omega/c$ (which is much less than $R_{\rm ej}$), then curvature radiation losses prevent acceleration to such high energies. For this reason, both \citet{Arons03} and \citet{Fang12} consider acceleration in the wind at radii $\gg R_{\rm LC}$ to avoid this problem.

Due to this difficulty, we instead consider the energy ions can be accelerated to just from confinement by magnetic fields within the nebula \citep{Lemoine15}, which gives
\be
	E_{\rm conf}(t) \approx ZeB_{\rm neb}(t)R_{\rm ej}(t).
\ee
The magnetic field that accelerates particles can also limit the energy of the UHECRs due to synchrotron losses. Equating the synchrotron cooling time with the gyration time of the particles results in
\be
	E_{\rm synch}(t) = \frac{3}{2}\frac{(Am_pc^2)^2}{(Ze)^{3/2}B_{\rm neb}(t)^{1/2}}.
\ee
With this in mind, we consider the highest possible energy for UHECRs at any given time to be the minimum between $E_{\rm conf}$ and $E_{\rm synch}$.

This maximum energy is shown in the upper panel of Figure \ref{fig:timescales} for UHECRs with either proton or Fe composition. At early times the nebula is dense enough in magnetic fields that synchrotron cooling dominates and the energy is $\approx E_{\rm sync}$. Then, as the nebula expands and the magnetar spins down, synchrotron losses decrease until the particle energy is instead $\approx E_{\rm conf}$. The maximum energy for UHECRs from a given magnetar is where these two energies are roughly equal. Also note that heavier nuclei generally reach higher energies because their larger charge promotes more acceleration.

Even though particles may be accelerated to such high energies, interaction with the environment around the magnetar may prevent the UHECRs from leaving unhindered. We now consider each of these, both in the ejecta and the nebula, in more detail.

\subsection{Interactions with the Ejecta}
\label{sec:ejecta}

For the ejecta, we use a hadron interaction model to estimate the timescale for sapping the UHECRs of their energy. This timescale is given by
\be
	t_{\rm ej}(E) = \lb c n_{\rm ej}\sigma_{\rm ej}(E)\xi(E) \rb^{-1},
\ee
where $\sigma_{\rm ej}$ is the cross section for hadronic interactions, and $\xi$ is the elasticity. Also note that all these terms will have energy dependencies in detail. The cross section for iron-proton interactions is $1.25\,{\rm barn}$, which applies when $Z=1$. Since the cross section should scale as roughly $Z^{2/3}$, for iron-iron interactions we estimate the cross section as $\approx11\,{\rm barn}$. Although $\xi$ can change significantly depending on the number of nucleons, we take it to be of order unity given the uncertainties in the cross sections which are degenerate with this. The relevant timescale to compare these to is the crossing timescale for the ejecta.
\be
	t_{\rm cross}(t) = R_{\rm ej}(t)/c.
\ee
When $t_{\rm cross} < t_{\rm ej}$, the UHECRs can escape the ejecta without significant interaction (and vice versa).

\begin{figure}
\epsscale{1.2}
\plotone{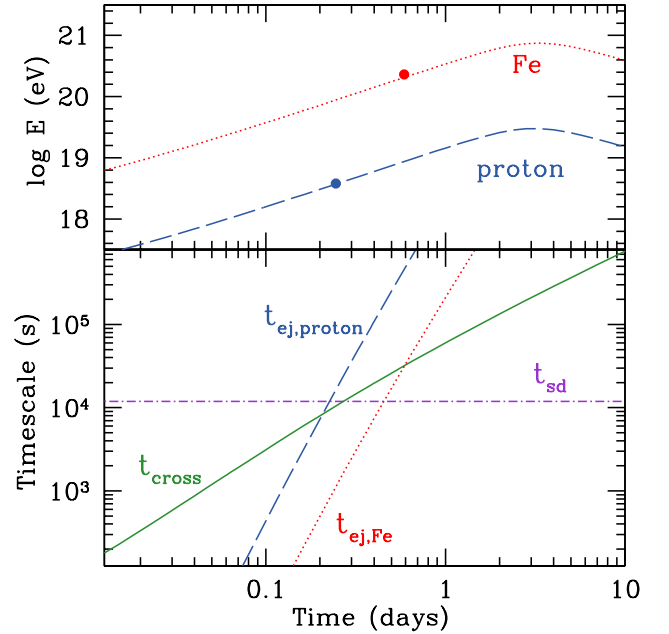}
\caption{In the upper panel we summarize the maximum energy the UHECRs can be accelerated to as a function of time from a newly born magnetar, using  $P_i=3\,{\rm ms}$, $B_*=5\times10^{14}\,{\rm G}$, and $M_{\rm ej}=0.01\,M_\odot$ for this example. The solid circles show where the UHECRs can begin escaping from the ejecta. The bottom panel summaries the evolution of relevant timescales important for understanding the escape of UHECRs from the ejecta surrounding the magnetar. The timescale $t_{\rm cross}=R_{\rm ej}/c$ (green, solid line) shows how long it takes a UHECR is escape the ejecta. This must be shorter than the hadron interaction timescale for escape to occur. These timescales are shown for both a proton-composition (blue, dashed line) and iron-composition (red, dotted line) UHECRs. Also plotted is the pulsar spin down timescale (purple, dot-dashed line), which shows when the energy output of the pulsar begins decreasing.}
\label{fig:timescales}
\epsscale{1.0}
\end{figure}

In Figure \ref{fig:timescales}, we present an example calculation of the relevant timescales for assessing the interaction of UHECRs with the ejecta along with maximum possible energy for the generated UHECRs as a function of time. We also compare different compositions of UHECRs, both protons (blue, dashed lines) and iron (red, dotted lines). This demonstrates that protons satisfy $t_{\rm cross}<t_{\rm ej}$ earlier during the evolution and  begin escaping earlier (everything to the right of the blue circle escapes). Nevertheless, comparing with the energy the protons would have at that time, protons do not reach as high of energies as the iron. Thus, if all other things are equal, a large charge is favored for producing the highest energy \mbox{UHECRs}.

\begin{figure}
\epsscale{1.2}
\plotone{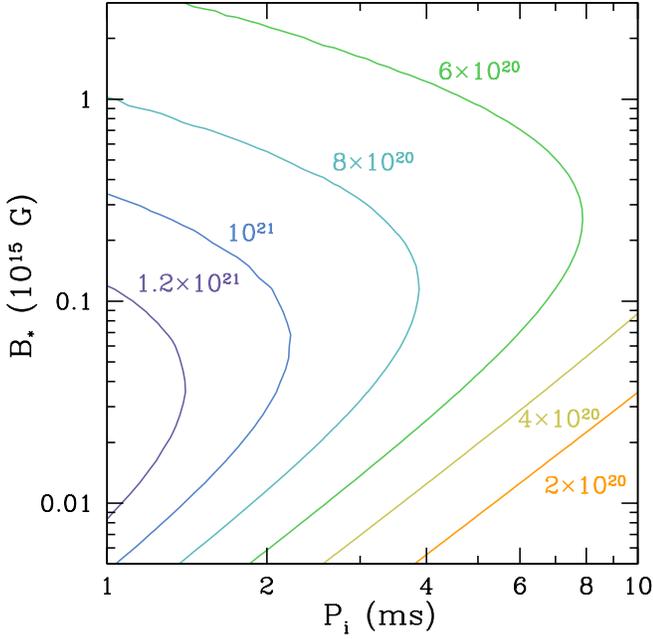}
\caption{Contours show the maximum possible energy for iron-composition UHECRs as a function of the initial spin period $P_i$ and magnetic field $B_*$ of the magnetar, using an ejecta mass of $0.01\,M_\odot$. Each contour is labeled in units of eV.}
\label{fig:contour}
\epsscale{1.0}
\end{figure}

With this picture in mind, we next estimate the maximum energy possible for iron-composition UHECRs as a function of the initial spin period $P_i$ and the magnetic field $B_*$. This is done by considering the highest possible energy for the UHECR past the time when they start escaping the ejecta (past the solid circles from Figure \ref{fig:timescales}). The results are summarized in the contour plot shown in Figure~\ref{fig:contour} where we use $M_{\rm ej}=0.01\,M_\odot$. For these calculations we assume $f=1$, which gives light curve solution that qualitatively match \citet{Metzger14} and allow us to cover a wide parameter space of models. In the future, a more detailed treatment of the thermalization should be conducted. This demonstrates that to produce the highest possible energy UHECRs requires a spin period of $P_i\approx2-4\,{\rm ms}$ and a rather modest (for a magnetar) magnetic field of $B\approx1-4\times10^{14}\,{\rm G}$ (note that although we consider $P_i\lesssim 2\,{\rm ms}$ unlikely for the reasons mentioned above, we include this parameter space in our plots for completeness). The reason why the highest field magnetars are disfavored for producing high energy particles is that they spin down too fast. Even though they produce high energy particles, these all interact with the ejecta at early times before it can become sufficiently diffuse.

When we consider more massive ejecta the maximum energies for the UHECRs decrease by slightly ($\sim10\%$ for $0.05\,M_\odot$) and the ``sweet spot'' for most efficiently accelerating UHECRs shift to slightly lower magnetic fields. This is because a lower magnetic field spins does the magnetar more slowly so that the time scale for the energy injection better matches the longer diffusion time through the more massive ejecta. With very high ejecta masses, as in the more associated with core-collapse \citep[e.g.,][]{Fang12}, the shift to lower energies can in some cases make it difficult to produce UHECRs at all. Overall though, the ejecta mass impacts the energetics less than $B_*$ and $P_i$ for our model, and thus we focus on these two parameters for most of the rest of our study.

\subsection{Photo-hadronic interactions}

A fraction of the X-rays coming out of the nebula thermalize with the ejecta. This generates a bath of thermal photons that can lead to photodisintegrations of the iron nuclei when they attempt to pass through. To estimate when this should occur, we largely follow the methodology presented in \citet{Kotera15}.

For ejecta with temperature $T_{\rm ej}$, the photon energy peaks at roughly $\epsilon_{\rm th}\approx 2.8k_{\rm B}T_{\rm ej}\approx2.4T_4\,{\rm eV}$, where $T_4=T_{\rm ej}/10^4\,{\rm K}$ (see Figure \ref{fig:properties2}). In contrast, the threshold energy of the photodisintegration process for iron-like nuclei is $\epsilon_A\approx 18.3A_{56}^{-0.21}\,{\rm MeV}$ \citep{Puget76,Murase09b}, where $A_{56}=A/56$ denotes the atomic mass of the nuclei. The resulting threshold Lorentz factor for photodisintegration is
\be
	\gamma_{\rm thres} \approx \frac{\epsilon_A}{\epsilon_{\rm th}}\approx 7.7\times10^6A_{56}^{-0.21}T_4^{-1},
\ee
or an energy of $E\approx4\times10^{17}\,{\rm eV}$ for iron-like nuclei. Thus, this threshold is always easily met in our scenario.

Besides exceeding this Lorentz factor, the the photodisintegration must be sufficiently fast in comparison to the crossing timescale \citep{Murase14}. Using a cross section of $\sigma_A\approx 8\times10^{-26}A_{56}\,{\rm cm^2}$ for a Giant Dipole Resonance \citep{Puget76,Murase08}, this timescale is
\be
	t_{\rm th} \sim
	\lp
		\frac{\sigma_AacT^4_{\rm th}}{\epsilon_{\rm th}}
	\rp^{-1}
	\sim 21 A_{56}^{-1}T_5^{-3}\,{\rm s},
	\label{eq:tth}
\ee
and thus strongly increases as the ejecta expands and cools. Nevertheless, during the first $\sim1-5\,{\rm days}$ of the evolution we generally find that $t_{\rm th}\ll t_{\rm cross}$ and thus the UHECRs will be subject to strong photo-hadronic interactions. This does not prevent iron UHECRs, but these estimates imply that lower mass nuclei will also necessarily be present. We save detailed calculations of the composition of the UHECRs as a function of energy for future work. We do note that once $t_{\rm th}\gtrsim t_{\rm cross}$, it is still possible that $E\gtrsim10^{20}\,{\rm eV}$ and thus UHECRs from iron-like nuclei will survive.

Another potential location for strong photo-hadronic interactions is within the nebula itself, which is full of non-thermal X-rays. Here the high energy of the individual photons make it easier to exceed the Lorentz threshold. Using the total energy in non-thermal X-rays, $E_{\rm nth}$, we thus estimate the timescale for interaction with these as
\be
	t_{\rm nth} \sim
		\lp
		\frac{\sigma_Ac}{\epsilon_{\rm nth}}
		\frac{3E_{\rm nth}}{4\pi R_{\rm ej}^3}
	\rp^{-1},
	\label{eq:tnth}
\ee
where $\epsilon_{\rm nth}$ is the energy of the photon. Thus, at higher photon energies this timescale is longer since the power-law distribution of photons is decreasing. This spectrum continues up to roughly the threshold for pair creation $\sim2m_ec^2\sim{\rm MeV}$.

\begin{figure}
\epsscale{1.2}
\plotone{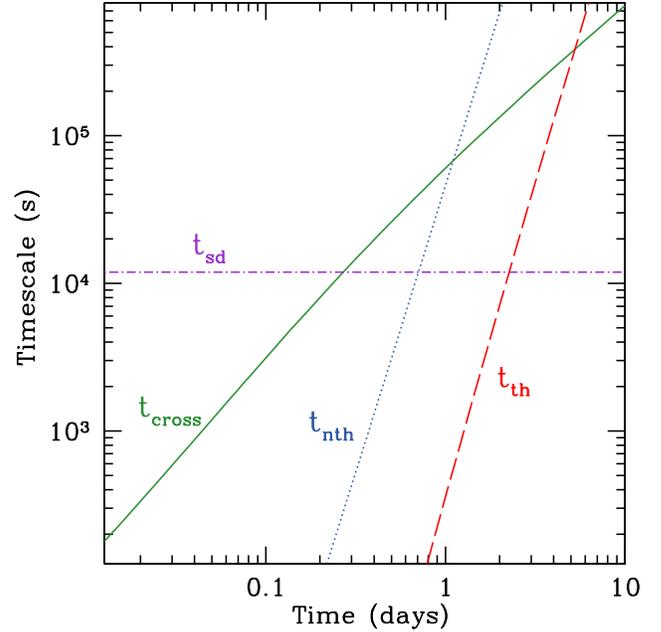}
\caption{Evolution of relevant timescales important for understanding the escape of UHECRs from the ejecta surrounding the newly born magnetar, using  $P_i=3\,{\rm ms}$, $B_*=5\times10^{14}\,{\rm G}$, and $M_{\rm ej}=0.01\,M_\odot$ for this example. The nebula crossing time $t_{\rm cross}$ and the spin down time $t_{\rm sd}$ are the same as in Figure \ref{fig:timescales}. The two additional timescales are that for photodisintegration within the nebula $t_{\rm nth}$ and with the ejecta $t_{\rm th}$, which are estimated in Equations (\ref{eq:tth}) and (\ref{eq:tnth}), respectively.}
\label{fig:timescales_hadronic}
\epsscale{1.0}
\end{figure}

We summarize the main arguments of this section in Figure \ref{fig:timescales_hadronic}, where we plot the relevant timescales for an example model. This demonstrates that where $t_{\rm th}<t_{\rm cross}$ photodisintegrations in the ejecta are relevant and this process generally dominates over the photodisintegrations within the nebula. This likely leads to a mixed composition of UHECRs up to a few days. After this timescale, the UHECRs can leave the magnetar less impacted by the environment.

\section{Global Energetics, Rates, and Constraints}
\label{sec:rates}

We showed in Section~\ref{sec:results} that the individual en caul magnetar is a potential site for UHECR generation.  We now investigate the expected energetics and rates associated with the population of such systems and the associated uncertainties therein.

\subsection{Order of Magnitude Estimates}

The energy generation rate of UHECRs above $10^{19.3}\,{\rm eV}$ is estimated to be in the range $\approx10^{43.5}-10^{44}\,{\rm erg\,Mpc^{-3}\,yr^{-1}}$ \citep{Waxman95,Berezinsky06,Katz09,Murase09}. In detail, such estimates depend on the assumed energy spectrum and volume over which the UHECRs can be observed, which may even be impacted by the (uncertain) composition. Nevertheless, such rough rates are helpful for constraining potential mechanisms. In the following discussions we use $10^{43.75}\,{\rm erg\,Mpc^{-3}\,yr^{-1}}$ as the characteristic rate to compare to for generating UHECRs.

For the magnetar model, the maximum energy available per event is $E_{\rm rot}$, but we do not expect this entire energy reservoir to be put into UHECRs.  Assuming a fraction $f_{\rm cr}$ does make it into UHECRs, the required rate, estimated from Equation~(\ref{eq:erot}) and the volumetric estimate above, is simply
\be
	R \approx 1.1\times10^{-8} I_{45}^{-1}\lp\frac{P_i}{2\,{\rm ms}} \rp^{2} f_{\rm cr}^{-1} 
	{\rm Mpc^{-3}\,yr^{-1}}.
	\label{eq:required_rate}
\ee
For AICs, it is helpful to compare to the Type Ia supernova rate since both types of events are produced from WDs. The Lick Observatory Supernova Search found a rate of $R_{\rm Ia}=(3.01\pm0.062)\times10^{-5}\,{\rm Mpc^{-3}\,yr^{-1}}$ \citep{Li11}, and thus, $R/R_{\rm Ia}\approx 4\times10^{-4} f_{\rm cr}^{-1}$.  This can be compared to the AIC rates that have been proposed in the literature, although these are known to be very uncertain. Using population synthesis, \citet{Yungelson98} find AIC rates of $8\times10^{-7}-8\times10^{-5}\,{\rm yr^{-1}}$ for the Milky Way, depending on assumptions about the common-envelope phase and mass transfer, which corresponds to  volumetric rate of $R_{\rm AIC}\sim 3\times10^{-9}-6\times10^{-7}\,{\rm Mpc^{-3}\,yr^{-1}}$. Similar constraints are also obtained from observed abundances of neutron-rich isotopes \citep{Hartmann85,Fryer99}.  This would give a relative rate  $R_{\rm AIC}/R_{\rm Ia}\sim 1\times 10^{-4} - 2\times 10^{-2}$, which is consistent with our model for values of $f_{\rm cr} \sim 0.01-0.1$.

The other scenario for making UHECRs from en caul magnetars we are considering is if some NS mergers produce massive NSs rather than BHs. Indeed, there is now evidence that some NSs could have masses well above $2\,M_\odot$ \citep{VanKerkwijk10,Romani12}, which favors an equation of state that makes such a scenario more likely. Both population synthesis and just the observed Galactic NS binaries favor a rate that is comparable to or greater than the AIC rate. For example, \citet{Kim05} estimate a Galactic rate of $\sim(0.1-3)\times10^{-4}\,{\rm yr^{-1}}$ or a volume rate of $\sim4\times10^{-8}-2\times10^{-6}\,{\rm Mpc^{-3}\,yr^{-1}}$. Again, this rate is consistent with what is needed for UHECRs for values of $f_{cr} \sim 0.01$, taxing neither the NS merger rate nor the available energy reservoir of the en caul magnetar population. 

\subsection{Detailed Energy and Rate Constraints}

The consistency arguments above imply a fraction of rotational energy $f_{\rm cr}\approx 0.01$ goes into UHECRs.  These rates are so uncertain, however, it is reasonable to ask whether such a value can be understood from independent lines of argument.  We now show that indeed the order-of-magnitude consistency arguments above agree nicely with more careful calculations.

\begin{figure}
\epsscale{1.2}
\plotone{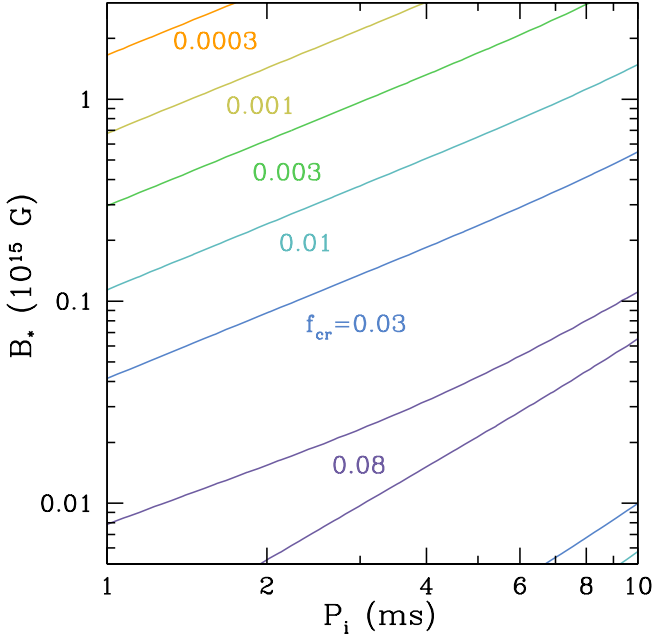}
\caption{The fraction $f_{\rm cr}$ of the rotational energy $E_{\rm rot}$ a given magnetar (or of the entire population of magnetars given a characteristic magnetic field and initial spin) that can put into UHECRs with $E>2\times10^{19}\,{\rm eV}$, assuming a fraction, \mbox{$\eta_{\rm cr}=0.1$,} of the spin down energy goes into UHECR acceleration.}
\label{fig:efficiency}
\epsscale{1.0}
\end{figure}

To understand how representative this value of $f_{\rm cr}$ is requires a slightly more detailed model of the UHECR production rate. To perform this, we first assume that the spectrum of UHECRs accelerated at any given time follows a $dN/dE\propto 1/E$ scaling, similar to previous discussions about pulsar magnetospheres  \citep{Arons03}. This assumes a Goldreich-Julian injection \citep{Goldreich69} and has mainly been chosen to illustrate the energy constraints on our model. In general, the spectrum could be different depending on the acceleration type \citep[for example, shock acceleration,][]{Lemoine15}, and in addition, other factor such as the details of the source population and the magnetic field within the nebula may also soften the spectrum \citep[e.g.,][]{Fang13}  Next, we assume that a fraction $\eta_{\rm cr}\lesssim1$ of the magnetar spin down can be used for UHECR acceleration. Energy conservation then requires
\be
	\int_{\rm E_{\rm min}}^{E_{\rm conf}} \frac{d\dot{N}}{dE} EdE \approx \eta_{\rm cr} L_{\rm sd},
	\label{eq:energy_conservation}
\ee
where $E_{\rm min}$ is the smallest energy for accelerated particles and can be ignored as long as $E_{\rm min}\ll E_{\rm conf}$. 
Equation (\ref{eq:energy_conservation}) can be solved to find
\be
	\frac{d\dot{N}}{dE} \approx \frac{\eta_{\rm cr}L_{\rm sd}}{E_{\rm conf}}\frac{1}{E}.
\ee
This describes the UHECRs produced within the nebula, but as we've discussed previously, they might not all escape the nebula. This is especially true at the highest energies due to synchrotron loses. To capture this physics, we only integrate $d\dot{N}/dE$ up to $E_{\rm synch}$ when $E_{\rm synch}<E_{\rm conf}$. The time-dependent energy output is then
\be
	L_{\rm cr}(t) &=& \left\{
              \begin{array}{cc}
		\eta_{\rm cr}L_{\rm sd}(t), \hspace{0.2cm} & E_{\rm synch}>E_{\rm conf}\\
           	\eta_{\rm cr}L_{\rm sd}(t)\lp\frac{E_{\rm synch}(t)}{E_{\rm conf}(t)}\rp, \hspace{0.2cm} & E_{\rm synch}<E_{\rm conf}
              \end{array}
       \right.
       \label{eq:lcr}
\ee
This is then time integrated over the evolution of the spinning down magnetar to get the total energy output in UHECRs. This is summarized in Figure \ref{fig:efficiency}, where we plot contours of the fraction $f_{\rm cr}$ of $E_{\rm rot}$ that goes into UHECRs. This calculation uses $\eta_{\rm cr}=0.1$, which is not dissimilar to the fraction of pulsar power found to experience the full potential drop by \citet{Chen14}, and these results can be simply scaled to other values of $\eta_{\rm cr}$ because the dependence is just linear. This shows that $f_{\rm cr}$ is greater than $\approx0.01$ over a significant fraction of the parameter space.

\begin{figure}
\epsscale{1.2}
\plotone{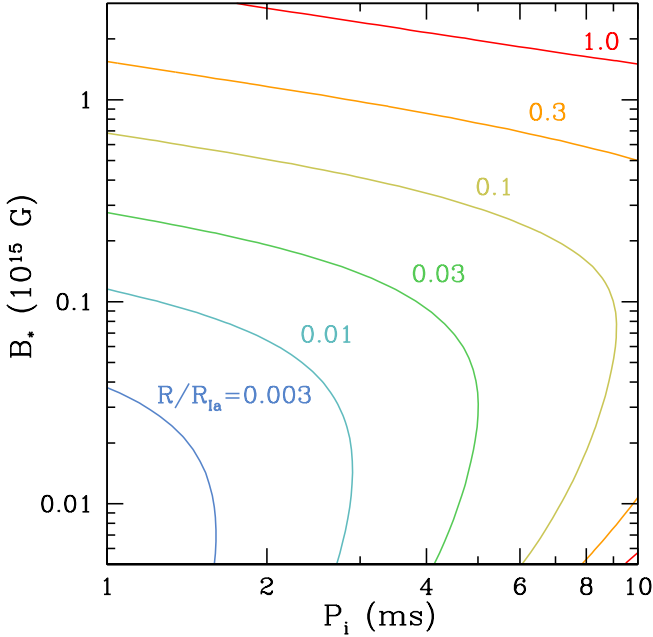}
\caption{Assuming that a given $B_*$ and $P_i$ are representative of most fast-spinning magnetars generated, the contours show the rate of events needed to explain the rate of UHECRs observed in units of the Type Ia supernova rate, taken to be $3.01\times10^{-5}\,{\rm Mpc^{-3}\,yr^{-1}}$ \citep{Li11}.}
\label{fig:sn_rate}
\epsscale{1.0}
\end{figure}

A further comparison is shown in Figure \ref{fig:sn_rate}. Assuming that a given $B_*$ and $P_i$ are representative of the majority of magnetars that produce UHECRs, we estimate what rate of events are needed to produce the observed UHECR rate, in units of the Type Ia supernova rate. As discussed above, typical AIC and neutron star merger rates are $\lesssim0.01$ in these units, thus this scenario favors the lower left quadrant of Figure \ref{fig:sn_rate}.

\subsection{Rate and Counterparts within the GZK volume}

UHECRs propagating through intergalactic space interact with the cosmic microwave background (CMB) or even other sources of extragalactic background light (EBL), which limits how far we can detect them.  This is known as the GZK effect \citep{Greisen66,Zatsepin66}. For both protons and Fe-like nuclei, the typical scale is similar and on the order of $D_{\rm GZK}\sim170\,{\rm Mpc}$ \citep{Metzger11}. Using the required rates estimated in Equation (\ref{eq:required_rate}), we then estimate
\be
	R_{\rm GZK} \approx 23 I_{45}^{-1}
	\lp\frac{P_i}{2\,{\rm ms}} \rp^{2}
	\lp\frac{f_{\rm cr}}{0.01} \rp^{-1}
	\lp\frac{D_{\rm GZK}}{170\,{\rm Mpc}} \rp^3\,{\rm yr^{-1}},
	\nonumber
	\\
	\label{eq:rate}
\ee
as the rate of events within the GZK volume. Such a distance scale is in fact interesting for other astrophysical interests. For example, this is similar to the horizon expected for detected NS mergers by ground-based interferometric detectors such as Advanced LIGO \citep{Abadie10}. This will either constrain or provide a direct measurement of NS merging events within the GZK volume.

In addition, the apparent bolometric magnitude of the electromagnetic transient of such an UHECR producing event would be $\approx15.5$ \citep{Metzger14} at $200\,{\rm Mpc}$. In principle, these events could have been detected by wide-field, transient surveys like the Palomar Transient Factory \citep[PTF;][]{Rau09} and the Panoramic Survey Telescope and Rapid Response System \citep[Pan-STARRS;][]{Kaiser02}. Unfortunately, given their particularly short timescales of $\sim1\,{\rm day}$ and blue color, it is possible they have so far been missed. Also, note that the events best suited for producing UHECRs will have smaller magnetic fields and correspondingly fainter electromagnetic counterparts.

In the future, this will change with surveys that have particularly rapid cadences, such as the Zwicky Transient Facility \citep[ZTF;][]{Law09}, the All-Sky Automated Survey for Supernovae \citep[ASAS-SN;][]{Shappee14}, or the Large Synoptic Survey Telescope \citep[][depending on the final cadence]{LSST}. In fact, such an event may have already been observed in GRB 080503 \citep{Perley09}, which showed a rebrightening in both optical and X-rays. Thus, perhaps another way to constrain the rate of these events is through GRBs (although the X-ray/optical signature is expected to be more isotropic than the beamed gamma-rays). Associated radio, either from the plerion associated with the magnetar \citep{Piro13} or interaction with the interstellar medium \citep{Nakar11} may also provide constraints. Either way, it appears that these events may be observable by independent means  on multiple different fronts if they are indeed making UHECRs \citep[for example, see][]{Metzger15,Horesh16}.  Within the next $\sim3-4\,{\rm yrs}$ we will have a much better empirical handle on the rates of such events and their role in UHECR acceleration.

\subsection{Intergalactic Propagation}
\label{sec:propagation}

UHECRs with large charges can be significantly deflected by the intergalactic magnetic field, so it is a good check to consider this effect in the context of our model. As a particle with charge $Ze$ travels over a magnetic field with correlation length $\lambda$, it typically gets deflected by an angle $\alpha\sim d/r_{\rm L}$, where $r_{\rm L} = E/ZeB_{\rm IGM}$ is the Larmor radius and $B_{\rm IGM}$ is the intergalactic medium  magnetic field. Traveling over a distance of $D$, the typical total deflection will be
\be
	\theta \sim \left( \frac{D}{\lambda}\right)^{1/2} \frac{\lambda}{r_{\rm L}}.
\ee
This deflection needs to be less than $\sim1\,{\rm rad}$ within the GZK volume, otherwise the path length of an UHECR will be increased by too much. This would decrease the effective GZK volume to the point that it may cause problems with getting a sufficient rate of events.  This limits the magnetic field to be
\be
	B_{\rm IGM} \lesssim 8\times10^{-9}Z^{-1}E_{20} \lp\frac{D_{\rm GZK}}{170\,{\rm Mpc}} \rp^{-1/2}
	\nonumber
	\\
	\times
	\lp\frac{\lambda}{1\,{\rm Mpc}} \rp^{-1/2}\,{\rm G},
	\label{eq:upperlimit}
\ee
where $E_{20} = E/10^{20}\,{\rm eV}$.

Another consideration is that the intergalactic magnetic field should provide enough smoothing such that at least one source within the GZK volume is providing UHECRs at any given time, otherwise there would be large angular anisotropies observed in the UHECR rate. For this to be satisfied, the diffusion time, which is roughly $\tau \sim \theta^2D/c $ should be longer than $R_{\rm GZK}^{-1}$, which we derived in Equation~(\ref{eq:rate}) above. This yields
\be
	B_{\rm IGM} &\gtrsim& 7\times 10^{-14} Z^{-1}E_{20}I_{45}^{1/2}
		\lp\frac{P_i}{2\,{\rm ms}} \rp^{-1}
	\lp\frac{f_{\rm cr}}{0.01} \rp^{1/2}
		\nonumber
	\\
	&&\times
	\lp\frac{D_{\rm GZK}}{170\,{\rm Mpc}} \rp^{-5/2}
	\lp\frac{\lambda}{1\,{\rm Mpc}} \rp^{-1/2}\,{\rm G},
	\label{eq:lowerlimit}
\ee
as a lower limit for the intergalactic magnetic field. Interestingly, these upper and lower limits show that in principle measuring the properties of the intergalactic magnetic field could help constrain the properties of the magnetars generating the UHECRs, albeit the wide range of the constraints probably make this difficult in practice.

\subsection{Galactic Propagation}

One is finally left wondering, ``What if an AIC occurred in the Milky Way?" Similar to the estimates made above, the propagation time across the disk of the Galaxy is
\be
	\tau \sim 58 Z^2E_{20}^{-2}
	 \lp \frac{D}{15\,{\rm kpc}} \rp^2
	\lp\frac{\lambda}{0.1\,{\rm kpc}} \rp
	\nonumber
	\\
	\times
	\left( \frac{B_{\rm ISM}}{3\times10^{-6}\,{\rm G}}\right)^2{\rm yr}.
\ee
For $Z=26$, this implies $\tau\sim4\times10^4\,{\rm yr}$ and thus for heavy nuclei the diffusion time is longer than the timescale between AIC or NS merger events in the Galaxy. So a non-zero fraction of UHECRs we measure today could have originated from an event in the Milky Way. The deflection angle is
\be
	\theta &\sim& 5\times10^{-2} ZE_{20}^{-1}
		 \lp \frac{D}{15\,{\rm kpc}} \rp^{1/2}
		 \nonumber
		 \\
	&&\times\lp\frac{\lambda}{0.1\,{\rm kpc}} \rp^{1/2}
	\left( \frac{B_{\rm ISM}}{3\times10^{-6}\,{\rm G}}\right){\rm rad},
\ee
which for $Z=26$ this implies deflection angles of $\sim50^\circ$, which is obviously not in the small angle approximation from which this formula was derived! The conclusion is that either $Z\sim1$ and there is very little contribution from the Galaxy or that $Z=26$ and although there is some contribution from the Galaxy, associating them with locations in the Milky Way will be difficult due to their large deflections from their origins.

\subsection{Constraints from High-Energy Neutrinos}

\begin{figure}
\epsscale{1.2}
\plotone{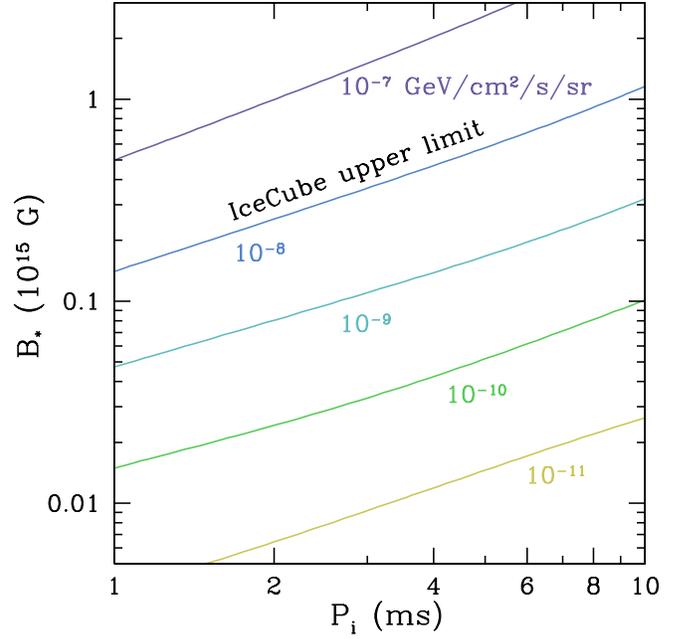}
\caption{Contours show the background flux of high-energy neutrinos expected from Equation (\ref{eq:neutrino_flux}) in units of ${\rm GeV\,cm^{-2}\,s^{-1}\,sr^{-1}}$. The rate $R$ is assumed to be sufficient to produce UHECRs with a rate of $10^{43.75}\,{\rm erg\,Mpc^{-3}\,yr^{-1}}$, and this assumes a volume of high-energy neutrinos out to a distance of $4000\,{\rm Mpc}$. The IceCube upper limit is resolved as function of energy, but is generally in the range of  $\sim10^{-8}\,{\rm GeV\,cm^{-2}\,s^{-1}sr^{-1}}$ or much higher.}
\label{fig:neutrinos}
\epsscale{1.0}
\end{figure}

Hadronic interactions between the UHECRs and the ejecta surrounding the magnetar will lead to charged pion production, which subsequently decay into high-energy neutrinos via $\pi^{\pm}\rightarrow e^\pm+\nu_e(\bar{\nu}_e)+\nu_\mu+\bar{\nu}_\mu$. Although a detailed study of this process is outside the scope of this paper, we make a simple estimate of the expected neutrino background to check for consistency with constraints from the IceCube Observatory \citep{IceCube,Aartsen13}.

The optical depth of UHECRs to hadron interactions, which we estimated in Section \ref{sec:ejecta} in the context of preventing UHECRs from escaping, is $\sim t_{\rm ej}/t_{\rm cross}$. Thus we estimate the efficiency of pion production as
\be
	f_\pi \sim \min( t_{\rm ej}/t_{\rm cross},1).
\ee
Furthermore, we assume that $0.2$ of the energy of a given UHECR can be converted to the energy of a pion, and $0.25$ of the pion energy then goes into the neutrino \citep[as was done in][]{Fang16}. This gives an overall conversion factor of $f_\nu\sim0.05$ from a given UHECR to a neutrino.

Integrating over the entire UHECR energy spectrum, we can derive the total energy of neutrinos that are generated. Note that in this particular case, $f_\pi$ can be taken outside of the energy integral since we are using a simplistic model for the hadronic interactions, which may not be true in a more detailed treatment. This greatly simplifies the evaluation of the integral, and total neutrino luminosity comes out to
\be
	L_{\nu}(t) \sim f_\nu f_\pi(t) \eta_{\rm cr}L_{\rm sd}(t).
\ee
Note that here we are using the {\it entire} luminosity that goes into UHECRs, which is represented by the term $\eta_{\rm cr}L_{\rm sd}$, rather than just the fraction that escape the magnetar environment, which is given in Equation~(\ref{eq:lcr}). This is because it is this luminosity of UHECRs that is important for generating the high-energy neutrinos. To relate this to the observed high-energy neutrino flux, we integrate $L_\nu$ over all time for a single event and over a distance $D$, resulting in
\be
	F_\nu = \frac{DR}{4\pi} \int_0^\infty L_{\nu}(t) dt,
	\label{eq:neutrino_flux}
\ee
where $R$ is the volumetric rate of events. We do not include additional neutrinos produced by secondary nuclei and pions, since our ejecta is so much less massive than in supernova-related scenarios \citep{Murase09b}.

The calculation of this high-energy neutrino background is summarized in Figure \ref{fig:neutrinos}. At each location in this plot we assume that $B_*$ and $P_i$ are representative of all UHECR-producing magnetars. Furthermore, we set $R$ to be the rate needed to match the observed UHECR rate (as summarized in Figure \ref{fig:sn_rate}). Finally we use $D=4000\,{\rm Mpc}$, essentially assuming a constant rate of UHECR production over the entire history of the Universe. A more detailed treatment that includes the time-dependent cosmic star formation rate and connecting this to the rate of en caul magnetar births over time is outside the scope of this work.

The IceCube upper limits are resolved as a function of energy and reach a minimum of $\sim10^{-8}\,{\rm GeV\,cm^{-2}\,s^{-1}sr^{-1}}$ at a neutrino energy of $\sim10^{14}\,{\rm eV}$. Furthermore, going up to energies of $\sim10^{18}\, {\rm GeV}-10^{20}\,{\rm GeV}$, the upper limits are $\sim3\times10^{-8}-3\times10^{-7}\,{\rm GeV\,cm^{-2}\,s^{-1}sr^{-1}}$. Comparing to Figure \ref{fig:neutrinos}, this is above our expected background rate over most of the parameter space, and note that we are integrating over neutrinos of all energies and thus likely overestimating the potential background. The reason we are able to meet the neutrino constraints is that the low amount of ejecta naturally gives $t_{\rm ej}/t_{\rm cross}\ll 1$ over most of the time when UHECRs are produced in the greatest amounts. Thus, for the same reason we are able to efficiently produce UHECRs, we are also able to not overproduce high-energy neutrinos. We note that there may also be photo-hadronic interactions within the nebula that may also produce neutrinos \citep{Lemoine15,Fang16}. Since this process is stronger for protons than heavy nuclei, and our main focus here is on the heavy nuclei as UHECRs, we save such a calculation for future work where the spectrum of UHECRs and the non-thermal radiation is computed in more detail.

\section{Discussion and Conclusions}
\label{sec:conclusions}

In this study we considered the production of UHECRs by quickly spinning magnetars following the AIC of WDs or the merger of NSs -- processes that set up the generic initial conditions we have termed ``en caul birth" as the newly born magnetar has a small shroud of ejecta in its immediate environs. Using a simple model of the nebula and ejecta surrounding the newly born magnetar, we considered both the acceleration of \mbox{UHECRs} and their ability to escape their production site, which each imprint features on the emergent \mbox{UHECRs}. We also explored the required rates for this scenario as well as the propagation of these particles though the intergalactic medium or even the interstellar medium, especially for the case where the nuclei are heavy. Perhaps most excitingly, the volume over which these \mbox{UHECRs} can propagate to Earth is similar to the distances that these events can be detected via electromagnetic and gravitational wave signatures so that it is likely this scenario will be tested within the next $\sim3-5\,{\rm yrs}$.

Looking to the future, a key issue for UHECRs will be conclusively measuring whether the composition is indeed heavy and how it is changing as a function of particle energy.  A heavy composition is difficult to reconcile with AGN and most GRB scenarios, but an iron-rich composition is naturally expected for a NS source such as young pulsars \citep{Fang12} or the magnetars studied here. A protomagnetar jet could also give heavy nuclei \citep{Metzger11} and in fact could reach elements such as zirconium, tellurium, or platinum at the highest energies, the measurement of which would be strong support for this model.

An important extrinsic difference between the en caul magnetar scenario and most other proposed scenarios for UHECRs is the galaxy environment. AIC and NS mergers are expected to occur in both old and young stellar environments, and in fact this is well-known from short GRBs \citep{Fong13}. This is very different than young pulsars or the long GRBs associated with protomagnetar jets that would be in young, star forming regions. Thus evidence that UHECRs are being produced in old stellar environments would be strong support for the model presented here.


\acknowledgments
We thank Andrew MacFadyen for important discussions when we were developing initial ideas for this work. We thank John Beacom for suggesting a comparison with the high-energy neutrino background. We also thank Jonathan Arons, Kumiko Kotera, Brian Metzger, Kohta Murase, E. Sterl Phinney, and Todd Thompson for their helpful feedback on previous drafts.


\end{document}